\documentclass{elsart}

\newcommand{\per}{$\rm Pb(ClO_4)_2 \,$}
\usepackage{graphicx}
\journal{Nuclear Instruments and Methods A}

\begin{document}
%\bibstyle{numbers}

\begin{frontmatter}

\title{Lead Perchlorate as a Neutrino Detection Medium}
\author{M.K. Bacrania},
\author{P.J. Doe},
\author{S.R. Elliott\corauthref{cor1}},
\corauth[cor1]{Corresponding author (sre@u.washington.edu)}
\author{C. E. Paul\thanksref{label1}},
\thanks[label1]{Present address: Raytheon, 2200 East Imperial Highway, El Segundo, CA 90245}
\author{L.C. Stonehill}, and
\author{D.I. Will}
\address{Center for Experimental Nuclear Physics and Astrophysics,
University of Washington, Box 354290, Seattle, WA, 98195}

\begin{abstract}
Lead can be an ideal medium for the detection and study of neutrinos.
Such a detector may be realized through the use of a lead perchlorate
(\per) solution as a Cerenkov radiator. The basic physical properties
of lead perchlorate solution are given and preparation of the
solution for use in a Cerenkov detector is described.  Results from
investigations of light transmission in lead
perchlorate solutions are also presented.

\end{abstract}

\begin{keyword}
% keywords here, in the form: 
neutrino detection \sep supernovas \sep lead perchlorate \sep cerenkov detector
\PACS 29.40.Ka \sep 14.60.Pq \sep 97.60.Bw
\end{keyword}
\end{frontmatter}

% main text

\section{Introduction}
In the energy regime of 10-30 MeV the neutrino interaction cross section \cite{kl,fhm}
on lead is 2-3 orders of magnitude greater than that of carbon, a
common detector medium. For this reason there has been
interest in using lead to study neutrinos from supernovae \cite{elliott,omnis,cline,hargrove,boyd}
or in accelerator-based neutrino oscillation searches \cite{konaka}.

Neutrino interactions with lead may occur via either the charged current
(CC) or neutral current (NC) mechanisms,

\begin{center}
\begin{equation}
\begin{array}{lclr}
   \nu_e + {\rm ^{208}Pb} & \rightarrow &{\rm ^{208}Bi^*} + e^- & {\rm
   (CC)}\\ & & \downarrow & \\ & & {\rm ^{208-Y}Bi} +
   X\gamma + Yn & \\ \\
   \nu_x + {\rm ^{208}Pb} & \rightarrow &{\rm
   ^{208}Pb^*} + \nu_{x}^{'} & {\rm (NC)} \\ & &
   \downarrow & \\ & &{\rm ^{208-Y}Pb} + X\gamma + Yn & \\
\end{array}
\end{equation}
\end{center}

where $X$ and $Y$ are the number of $\gamma$ rays and neutrons emitted, respectively.
$Y$ can be either 0, 1, or 2, depending on the
incident neutrino energy and the resulting excited nuclear state.
 The energetics of these transitions are shown in
Fig. 1.

\begin{figure}%[p]
\label{fig:levelscheme}
\begin{center}
\includegraphics[width=4. in]{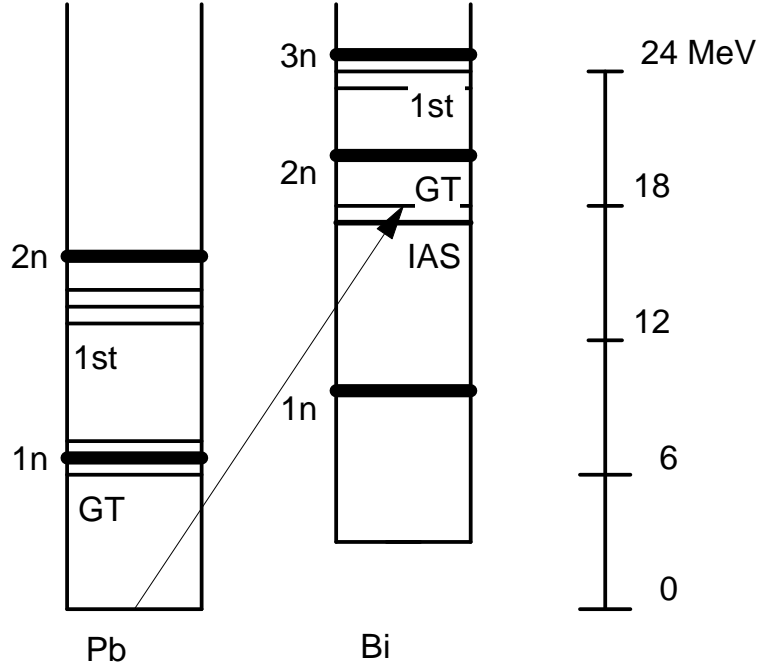}
\caption{The level scheme of \nuc{208}{Pb} - \nuc{208}{Bi} system. The levels are labeled
 {\it GT} to indicate the Gamow-Teller resonances, {\it IAS} to indicate the isobaric analog state, and
{\it 1st} to indicate the states populated by first forbidden transitions. The {\it 1} and {\it 2} neutron 
emmision thresholds are indicated by the labels 1n and 2n respectively.}
\end{center}
\end{figure}

Lead perchlorate (\per) has a very high solubility in water. A
saturated solution consists of approximately five parts by weight of
lead perchlorate to one part water (see Table 1) and is
transparent to the eye. These properties led to its early consideration
as a Cerenkov radiator \cite{jelly} and for gamma ray detection
\cite{cern} \cite{may}. The presence of hydrogen in the solution results
in efficient thermalization of any neutrons associated with neutrino
interactions. There is a high probability ($>$90\%) that these neutrons will then
capture on the $^{35}$Cl present in a saturated solution, with the subsequent
emmission of gamma rays totalling 8.6 MeV. As discussed by the Sudbury
Neutrino Observatory collaboration
\cite{sno}, these gamma rays may then be detected via their Compton
scattered electrons in a Cerenkov detector.

\begin{table}[ht]
\label{tab:proptable}
\begin{center}
\begin{tabular}{ll}
\hline
Property                                & Value \\ \hline \hline
$^{208}$Pb number density               & 1.19 x $10^{21}$ cm$^{-3}$ \\
H number density			& 4.37 x $10^{22}$ cm$^{-3}$ \\
$^{208}$Pb$(\nu_e, e^-)$ cross section at 30 MeV      & 34 x
$10^{-40}$ cm$^2$
\\
$^{35}$Cl n capture cross-section       & 44.0 b \\
Density                                 & 2.1 gm cm$^{-3}$     \\
Refractive index                        & 1.45\\
Attenuation Length                      & $\approx$4 m \\  \hline
\end{tabular}
\end{center}
\caption{Some properties of a 70\% lead perchlorate solution prepared in the manner 
described in the text.}
\end{table}

While both the CC and NC reactions result in neutron and gamma
production, only the CC reaction produces a prompt
electron. This is important since detection of the prompt electrons
allows the separation of CC and NC events.

The CC reactions may further be divided into $\nu_e$ and $\bar{\nu}_e$
events. The $\nu_e$ events in lead can
produce one or two neutrons. The $\bar{\nu}_e$ events result from
interactions with hydrogen in the solution, and only produce a single
neutron. Thus the two-neutron spectrum contains only $\nu_e$ events,
while the one-neutron spectrum contains both $\nu_e$ and $\bar{\nu}_e$
events. By comparing the two spectra the $\nu_e$ and $\bar{\nu}_e$
reactions can be separated. The reaction $\bar{\nu}_e$-Pb is
relatively insignificant due to its smaller cross-section \cite{kl,fhm}.

Neutrino interactions in lead that occur through the NC
process can produce, via a Gamow-Teller resonance, a single 7.6-MeV
gamma ray and no neutron. Alternatively, the NC reaction may result in
a single neutron, with little or no electromagnetic energy. 
Two-neutron production by the NC reaction has
a lower cross section than for the CC reaction
\cite{kl,fhm}.

Thus, by careful measurement of the prompt energy and the number of
neutrons produced, one may determine the neutrino flavor and, in the
case of CC interactions, the energy of the interacting neutrino.
Details of these analyses in \per\ are discussed along with their applications to
the study of supernovae in Ref. \cite{elliott}.

\section{Applications of a lead perchlorate based neutrino detector}

The interaction cross sections of neutrinos with complex nuclei are of great
importance in supernova studies. They are central to the supernova
explosion mechanism, the nucleosynthesis of heavy elements in
supernovae, and the detection of supernovae neutrinos. Therefore 
accurate values for the cross sections are
of both theoretical interest and an
essential ingredient in the design and interpretation of a lead based supernova
neutrino detector. Unfortunately, the
 theoretical predictions of the total inelastic neutral
current and charged current cross sections for neutrino reactions on
$\rm ^{208}Pb$ given by \cite{kl} and \cite{fhm} differ significantly and 
the cross sections for neutrino reactions on various
isotopes of lead have not been measured.   However because the neutrino energies are similar, a
 lead perchlorate based detector at a stopped-pion neutrino
source would be a very effective way of measuring the $\nu$-Pb cross
sections in the energy region of interest to supernova studies.
 For example, at the proposed ORLAND/SNS neutrino source
\cite{orland}, a 10 tonne \per detector could measure the $\nu_e$-Pb
cross section to an accuracy of 10\% with approximately 100 days of
detector livetime.

Lead offers promise as a supernova neutrino detection medium because the
neutron production from neutrino interactions with lead is extremely
sensitive to the energy spectrum of the supernova electron neutrinos.
This sensitivity is due to the pronounced increase in the $\nu_e$-Pb
cross section in the energy region significant to supernova neutrinos. In 
particular, the 2-neutron production cross section increases dramatically with energy.
The measured one- and two-neutron spectra from a Pb-based
detector and the 2 neutron/1 neutron event ratio provide data sensitive to any 
potential neutrino oscillation processes occurring in a supernova\cite{elliott}.
 In the absence of oscillations,
the expected energy hierarchy for neutrino production in supernovae is
$\bar{E}_{\nu{_e}} < \bar{E}_{\bar{\nu}{_e}}<\bar{E}_{\nu_{\mu,\tau}}$. A
\per neutrino detector could exploit this hierarchy to discriminate
between neutrino flavors. If higher-energy mu and tau neutrinos oscillate
into electron neutrinos, both the average detected electron neutrino
energy and the 2-neutron CC interaction rate would increase dramatically.
Observing this increase would be strong evidence that neutrino
oscillations are taking place. Fuller, {\it et al.} \cite{fhm} predict that the
definitive signal of tau to electron oscillations in a $\rm ^{208}Pb$
detector is a dramatic enhancement (up to a factor of 40) in multiple
neutron events. 

\section{Filtration and Attenuation Length Measurement}
The attenuation length of light in a material may be defined as the
length of material over which the intensity of light decreases by a
factor of 1/e.  For a solution to be considered for use in a Cerenkov
detector the attenuation length should be large compared to the
dimensions of the detector.  This sets the scale for detector dimension or
segmentation.

\begin{figure}%[p]
\label{fig:attenschem}
\begin{center}
\includegraphics*[width=7cm]{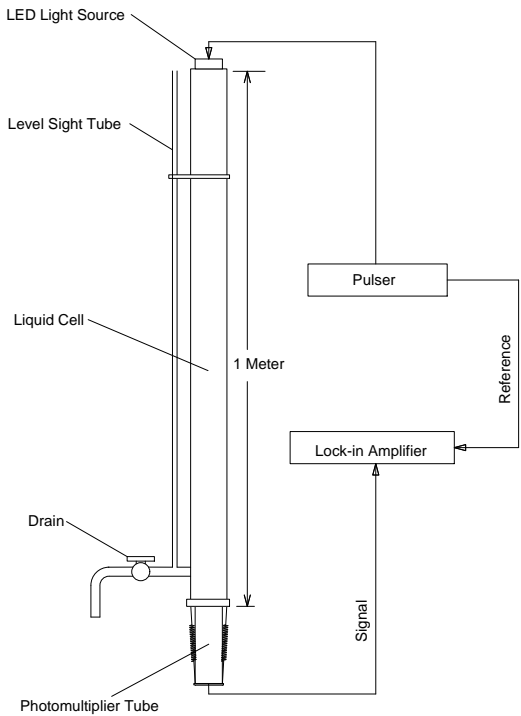}
\caption{A schematic drawing of the light attenuation measurement apparatus.}
\end{center}
\end{figure}

In order to assess the feasibility of using lead perchlorate as a
Cerenkov neutrino detection medium, we constructed an apparatus to
measure the attenuation length of various lead perchlorate solutions
(see Fig. 2). The apparatus consists of a 1.3-meter
tall, 3-cm inner diameter, chlorinated polyvinyl chloride (CPVC) column. A
430 nm light emitting diode (LED), powered by a 1-kHz square-wave, is
mounted at the top of the column. The light from the LED is focused onto
a 1.9 cm diameter, flat-faced (Hamamatsu R1450) photomultiplier tube
(PMT) mounted at the bottom of the column. The signals from the PMT,
along with the LED pulse-generating waveform, are fed into a lock-in
amplifier. To measure the attenuation length, the column is filled with
the lead perchlorate solution, and as the the liquid level is slowly
lowered, the output of the PMT is recorded. Plotting the PMT output as a
function of the liquid level $x$, yields an exponential curve $y =
Ce^{-x/\lambda}$, where
$\lambda$ is defined to be the attenuation length of the solution.

Lead perchlorate solutions were obtained from GFS
Chemicals \cite{gfs} in 50\% and 82.4\% (saturated) solutions by mass.
The saturated \per solution contained visible precipitates and had an
initial attenuation length of approximately 20 cm. The saturated
solution was diluted to 70\% concentration by first mixing it with
deionized water and heating it to $\rm 42^{\circ}C$ while stirring
overnight. This procedure removed the visible precpitates and
increased the attenuation length to 54 cm. In order to determine
whether any lead salts still present in the solution were affecting
the attenuation length, a means of filtering the solution was
developed. The filtration system consisted of a positive-displacement,
compressed-air-driven Warren Rupp Marathon Pump (model MP01P)
constructed from polyvinylidenedifloride (PVDF) and compatible with lead
perchlorate solutions. This pump was used to transfer solutions from one
reservior to another via a 47-mm diameter polypropylene filter holder
(Advantec MFS, Model 501200).  The primary consideration for filtering
lead perchlorate is the compatibility of the filter material with the
lead perchlorate solution. Polytetrafluoroethylene (PTFE) filters were
found to be the most compatible, but have the disadvantage of being
hydrophobic.  The recommended procedure for using PTFE filters is to wet
them with alcohol before filtrations. Unfortunately, alcohol and lead
perchlorate are incompatible, so wetting was impossible. Thus, the
usable pore size of unwetted PTFE filters was limited to 1.0 micron;
smaller pore sizes did not allow even pure water to pass through
without puncturing the filters.  PVDF filters were also quite
compatible as long as the pH of the lead perchlorate solution was
$>4$.  It is necessary to maintain the acidity of the solution to
minimize the formation of salts, but allowing the solution to become
too acidic resulted in the disintegration of the PVDF filters. PVDF has
the distinct advantage over PTFE of being hydrophilic, which allowed the
use of smaller pore sizes.

While \per-induced deterioration of the CPVC column was negligible, PVC
filters suffered deterioration.  Glass-fiber filters and ceramic (alumina)
filters were also found to be highly unsatisfactory, as both media seemed
to introduce contaminants into the solution, even though the ceramic did
not visibly deteriorate.

After each filtration, the attenuation length of the filtered solution
was measured.  The curve for a measurement was fit to an exponential
and the best fit value for $\lambda$ was taken as the attenuation
length for that measurement.  For some solutions this attenuation
length measurement was repeated two or three times and the reported
attenuation length and uncertainty is the average and spread of these
measurements. For other solutions the uncertainty was estimated based on
the characteristic uncertainties of the repeated measurements.

For each solution the refractive index was measured using a standard
refractometer. Each measurement was repeated a number of times and the spread 
in the measurements was very small ($\pm 0.0001$).  However a measurement of water and Isopropanol
found a difference of $\approx$ 0.003 with respect to the known values.
We represent the uncertainty with this latter larger value. The
density was measured using a
 10-ml glass gravity pycnometer to an accuracy
of $\pm$0.3\%. The index of refraction and the density both scale linearly
with the molarity of the solution.

The most successful sequence of filtrations started with a 70\% lead
perchlorate solution, which was filtered through a progressively
finer series of filters, with a single pass through each filter.  The pore
sizes and material were: 5.0 micron PTFE, 2.0 micron PTFE, 1.0 micron
PTFE, 0.2 micron PVDF.  This filtration sequence resulted in an
attenuation length of $422.5 \pm 14.5$ cm. The attenuation curve from one
measurement of this filtered solution is shown in Fig. 3.

\begin{figure}%[p]
\label{fig:attenuation}
\begin{center}
\includegraphics*[width=14cm]{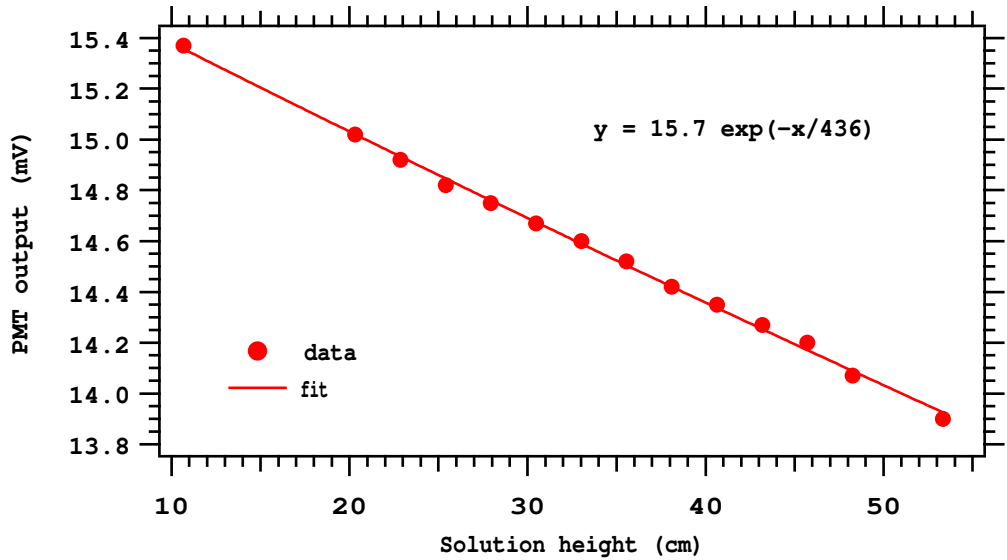}
\caption{An attenuation length curve for filtered 70\% \per\ solution obtained using 430 nm light.}
\end{center}
\end{figure}

The improvement in the attenuation length with each stage of the
filtering sequence is shown in Table 2. Also shown are
the measured values of the liquid density and refractive index after
each filtering step. The transmission spectrum for a 60\% \per solution which has undergone
an identical filtering sequence is shown in Fig. 4. 

\begin{table}[ht]
\label{tab:attentable}
\begin{center}
\begin{tabular}{lccc}
\hline
Filter type & Attenuation $(\rm cm)$ & Density $(\rm g\, cm^{-3})$ &
Refractive Index \\ \hline \hline
unfiltered           & $9.5 \pm 1.5$    & 2.164(6) & 1.455(3)\\
5.0 $\rm {\mu}m$ PTFE& $14.5 \pm 2.5$   & 2.119(6) & 1.450(3)\\
2.0 $\rm {\mu}m$ PTFE& $49.0 \pm 5.9$   & 2.122(6) & 1.450(3)\\
1.0 $\rm {\mu}m$ PTFE& $322.7 \pm 14.9$ & 2.114(6) & 1.449(3)\\
0.2 $\rm {\mu}m$ PVDF& $422.5 \pm 14.5$ & 2.098(6) & 1.448(3)\\ \hline
\end{tabular}
\caption{The filtration results for 70\% solution of lead perchlorate. The
attenuation length was measured using 430 nm light.}
\end{center}
\end{table}

\begin{figure}%[p]
\label{fig:transpec}
\begin{center}
\includegraphics*[width=10cm]{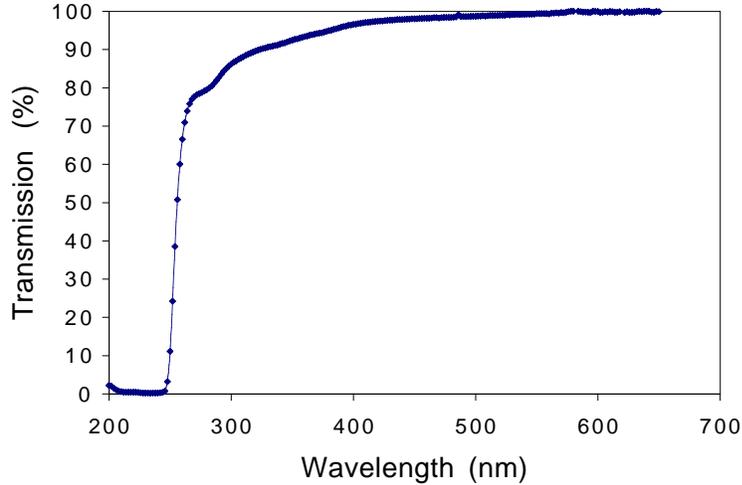}
\caption{The transmission spectrum with respect to pure water for a 60\% \per\ solution.}
\end{center}
\end{figure}

\section{Conclusions}
The neutrino continues to offer unique insights into both the standard
model of particle physics and the nature of extreme physical phenomena
such as supernovae. A Cerenkov detector using a solution of lead
percholorate has the ability to both measure neutrino energy as well
as discriminate between neutrino flavors, features which make it an
ideal detector to study neutrinos in the regime of 10-30 MeV. Simple
filtration techniques have yielded light attenuations lengths suitable
for accelerator neutrino studies and astrophysical neutrino
detectors.

\section{Acknowledgements}
We would like to thank Hank Simons and Jim Elms (UW CENPA), and Bob Morley
(UW Physics) for their technical expertise in constructing various parts
of the experimental apparatus. We also thank Jim Patterson and Jim Roe
from the UW Department of Chemistry for their assistance with measuring
the refractive indices (JP) and transmission spectra (JR) of the lead
perchlorate solution.

\end{document}